# Optimizing the Temporal and Spatial Resolutions and Light Throughput of Fresnel Incoherent Correlation Holography in the Framework of Coded Aperture Imaging

FRANCIS GRACY AROCKIARAJ,[1,2,†] AGNES PRISTY IGNATIUS XAVIER,[1,2,†] SHIVASUBRAMANIAN GOPINATH,[1,†] ARAVIND SIMON JOHN FRANCIS RAJESWARY[1], SAULIUS JUODKAZIS, [3,4] AND VIJAYAKUMAR ANAND[1,3,*]

[1]Institute of Physics, University of Tartu, W. Ostwaldi 1, 50411 Tartu, Estonia
[2]School of Electrical and Computer Engineering, Ben Gurion University of the Negev, P.O. Box 653, Beer-Sheva 8410501, Israel
[3]Optical Sciences Center and ARC Training Centre in Surface Engineering for Advanced Materials (SEAM), School of Science, Computing and Engineering Technologies, Swinburne University of Technology, Hawthorn, Melbourne, VIC 3122, Australia
[4]Tokyo Tech World Research Hub Initiative (WRHI), School of Materials and Chemical Technology, Tokyo Institute of Technology, 2-12-1, Ookayama, Meguro-ku, Tokyo 152-8550, Japan
[†]These authors contributed equally to this work
*vijayakumar.anand@ut.ee

**Abstract:** Fresnel incoherent correlation holography (FINCH) is a well-established digital holography technique for 3D imaging of objects illuminated by spatially incoherent light. FINCH has a higher lateral resolution of 1.5 times that of direct imaging systems with the same numerical aperture. However, the other imaging characteristics of FINCH such as axial resolution, temporal resolution, light throughput and signal to noise ratio (SNR) are lower than those of direct imaging system. Different techniques were developed by researchers around the world to improve the imaging characteristics of FINCH while retaining the inherent higher lateral resolution of FINCH. However, most of the solutions developed to improve FINCH presented additional challenges. In this study, we optimized FINCH in the framework of coded aperture imaging. Two recently developed computational methods such as transport of amplitude into phase based on Gerchberg Saxton algorithm (TAP-GSA) and Lucy-Richardson-Rosen algorithm were applied to improve light throughput and image reconstruction respectively. The above implementation improved the axial resolution, time resolution and SNR of FINCH close to those of direct imaging while retaining the high lateral resolution. A point spread function (PSF) engineering technique has been implemented to prevent the low lateral resolution problem associated with the PSF recorded using pinholes with a large diameter. We believe that the above developments are beyond the state-of-the-art of existing FINCH-scopes.

**Keywords**: Incoherent digital holography; Fresnel incoherent correlation holography; Lucy-Richardson-Rosen algorithm; 3D imaging; coded aperture imaging; Gerchberg-Saxton algorithm

## 1. Introduction

Fresnel incoherent correlation holography (FINCH) is a well-known incoherent digital holography technique [1,2]. FINCH was developed by Rosen and Brooker in 2007 for 3D imaging of incoherently illuminated objects. In FINCH, light from an object point is split into two and differently modulated using two diffractive lenses with different focal lengths and interfered to form the self-interference hologram. In the first version of FINCH, phase masks formed by a random multiplexing of two diffractive lens phase masks with three relative phase shifts $\theta = 0$, $2\pi/3$ and $4\pi/3$ were displayed on a spatial light modulator one after the other and the resulting three holograms were recorded. The recorded holograms were combined to create a complex hologram which was numerically backpropagated in the computer to reconstruct the object information at any plane of interest without twin image and bias terms. At the time of invention of FINCH in 2007, there was no existing incoherent 3D imaging technique that was as robust and simple as FINCH. In a way, FINCH created a new direction for modern day incoherent digital holography technologies. FINCH gained much attention after its demonstration on a fluorescence microscope [2] resulting in one of the highly researched topics and patented technologies in the area of incoherent imaging.

During subsequent research on FINCH, again by the same team of Rosen and Brooker, the super resolution capability of FINCH was revealed [3]. FINCH has a resolving power which is 2 times and 1.5 times that of coherent and incoherent imaging systems with the same numerical aperture (NA). The resolution enhancement in FINCH is unlike any other resolution enhancement methods such as structured illumination (SI) [4]. In FINCH the resolution enhancement is obtained by shaping the modulation transfer function (MTF) within the spatial frequency boundaries imposed by the NA, whereas in SI approach, the bandwidth of MTF is changed. In a direct incoherent imaging system, the MTF has a high response for low spatial frequencies and a low response for high spatial frequencies. In FINCH, at the optimal operating conditions, the MTF has a uniform response. With SI, when the MTF is expanded, the fundamental response structure of MTF of direct imaging is carried on preventing realization of the full potential of SI. Therefore, FINCH can be considered as a more fundamental imaging system with the optimal MTF that can be used as a platform for applying other super resolution technologies such as SI [5]. The unusual MTF of FINCH results in surprising imaging characteristics. The magnification of spacing between two points is twice as much as the magnification of a point when imaging using FINCH with an optimal optical configuration. Besides, the axial resolution of FINCH is lower than that of an incoherent imaging system with the same NA. When the super resolution capability of FINCH was demonstrated, the optical configuration of FINCH was also upgraded from spatial multiplexing with random mask to polarization multiplexing method which improved the signal to noise ratio (SNR) significantly. The evolution of FINCH is presented in details in [3, 6].

Starting from the version of FINCH demonstrated in [3], FINCH evolved rapidly over the past decade with contributions made by leading researchers around the world [6-8]. Before describing the contributions of different groups in FINCH, the state-of-the-art FINCH in 2012 is presented. The reported version of FINCH in [3] used polarization multiplexing approach and recorded three phase-shifted holograms with three camera shots. This version when compared to direct incoherent imaging systems with same NA exhibited a superior lateral resolution but an inferior axial resolution and an inferior temporal resolution of one-third. The light throughput in FINCH is about 25% that of other incoherent imaging systems. With implementation, FINCH required an active device like a spatial light modulator and numerous optical elements such as lenses and polarizers. Many developments were made in FINCH to reduce one or many of the above disadvantages out of which some key contributions are described next.

Several groups attempted to improve the temporal resolution of FINCH to match that of direct incoherent imaging systems. Spatial multiplexing methods were reported by Nobukawa and Nomura [9,10], polarization multiplexing method using a 4-pol image sensor where every pixel consists of 2×2 array of polarizers at 0º, 45º, 90º and 135º was reported by Tahara [11]. The above methods improved the temporal resolution of FINCH at the expense of field-of-view [9,10] and SNR [11]. Passive optical elements such as geometric lens [12] and metasurfaces [13] were manufactured to make FINCH compact and light-weight by Liu and Huang respectively. Different recording and reconstruction methods were reported by Tahara [14] and Anand [15]. While [14] required at least two camera shots for FINCH, [15] demonstrated single camera shot FINCH with a one-time recorded point spread function (PSF) and non-linear reconstruction (NLR) method [16]. The axial resolution of FINCH was improved by a scanning pinhole approach [17]. FINCH concept was adapted for other applications such as localization microscopy [18,19], light sheet microscopy [20] and advanced manufacturing [21].

FINCH in different configurations have different set of challenges. In the widely used configuration of imaging, FINCH has one-half of the temporal resolution [13], one-fourth of light throughput and lower axial resolution than other incoherent imaging systems. The implementation of FINCH requires either SLM or other liquid crystal element [22]. In all the above studies every solution came with a problem except for [14]. The study [15] was peculiar as FINCH was implemented as a coded aperture imaging method [23]. This is the only study where the axial resolution of FINCH was improved unexpectedly. A follow-up study of FINCH as coded aperture imaging confirmed the improvement in axial resolution of FINCH [24]. One drawback of [15] was that the random multiplexing method used for creating the phase masks and the NLR method resulted in a low SNR. Recently, advancements have been made on the above two fronts. A novel computational algorithm called Transport of Phase into Amplitude based on Gerchberg Saxton Algorithm (TAP-GSA) for spatially multiplexing two phase masks with minimum random noises and a high SNR reaching the limit of direct incoherent imaging was developed [25]. An advanced computational reconstruction method called Lucy-Richardson-Rosen algorithm (LR$^2$A) was developed as a better alternative to NLR and implemented in many studies for 3D imaging [26-28]. A preliminary study of FINCH using TAP-GSA and LR$^2$A was carried out which demonstrated a superior performance of FINCH [29]. One recent study of combining FINCH with another holography technique called coded aperture correlation holography (COACH) [30] to obtain the axial resolution of COACH and lateral resolution of FINCH [31]. This new technique is called coded aperture with FINCH intensity responses (CAFIR). However, CAFIR required three camera shots. A most recent study from our research group on self wavefront incoherent transverse splitting holography (SWITSH) a derivative of FINCH showed promising results of single shot with LR$^2$A.

In this study, we attempt to optimize FINCH on all the aspects of imaging using TAP-GSA, LR$^2$A and PSF engineering with all known advantages and devoid of the inherent disadvantages of FINCH. The optimized FINCH, has the same temporal resolution, similar axial resolution, same light throughput and SNR as a direct incoherent imaging system and retains the advantages such as the superior lateral resolution and 3D imaging capability. The proposed FINCH with the above collection of methods can be implemented with a passive diffractive element, does not require vibration isolation system and additional optical components. The PSF engineering approach allows to shift the lateral resolution limit from the size of the pinhole to that defined by the NA of the imaging system. The rest of the manuscript is organized as follows. The methodology is presented in the second section. The simulation studies of 2D and 3D imaging are presented in the third section. In the fourth section, experimental results are presented. The conclusion and future perspectives are presented in the final section.

## 2. Materials and Methods

The computational optical configuration of FINCH implemented as CAI system is shown in Figure 1. The imaging process consists of three major steps: mask design, optical recording and computational reconstruction. In the first step, the two phase-only diffractive lens masks are spatially multiplexed to form a single phase-only mask using TAP-GSA [25]. The designed mask is either manufactured using lithography techniques or displayed on a SLM to record the point spread hologram (PSH) in the first step and an object hologram (OH) in the next step under identical recording conditions [24]. In the third and final step, the recorded holograms are reconstructed using LR$^2$A with suitable conditions to obtain optimal reconstruction results [26]. Assuming that the distances are large, Fresnel approximation-based equations are used for analyzing the system. A point object emitting light with an amplitude of $\sqrt{I_o}$ is considered. The point object is located at a distance of $z_s$ from a refractive lens L as shown in Figure 1. To achieve the optimal FINCH condition of super-resolution, it is necessary to have the same beam diameter for the two beams at the sensor plane located at a distance of $z_h$ from the SLM. The elements L and SLM are assumed to be in tandem to simplify the analysis. To achieve the optimal FINCH condition, when the object distance $z_s$ equals the focal length $f$ of the refractive lens L, the focal length of the two diffractive lenses need to be $z_h/2$ and $\infty$ respectively. Only for the plane $z_s = f$, FINCH will exhibit an optimal resolution and has a lower resolution in other planes. The phase function of the two lenses are $\exp(-ikr^2/z_h)$ and 1, where $r = \sqrt{x^2 + y^2}$ and $k = 2\pi/\lambda$ and $\lambda$ is the wavelength. Using TAP-GSA, the two lenses are combined to form a phase-only diffractive element given as $\exp[ikP(x,y)] = 1 + \exp(-ikr^2/z_h)$, where $P(x,y)$ is the optical path length variation.

The complex amplitude from the point object reaching the lens L is $C_1\sqrt{I_o}Q(1/z_s)$, where $C_1$ is a complex constant, $Q$ is a quadratic phase function given as $Q(a) = \exp(ikar^2/2)$. The complex amplitude after the lens L is $C_2\sqrt{I_o}Q(1/z_s)Q(-1/f)$, and the complex amplitude after the SLM is $C_3\sqrt{I_o}Q(1/z_s)Q(-1/f)\exp[ikP(x,y)]$, where $C_2$ and $C_3$ are complex constants. The PSH recorded at the sensor plane is given as

$$I_{PSH} = \left|C_3\sqrt{I_o}Q(1/z_s)Q(-1/f)\exp[ikP(x,y)]\otimes Q(1/z_h)\right|^2, \quad (1)$$

where '$\otimes$' is a 2D convolutional operator. If $O$ is a multipoint object, then the object hologram (OH) can be given as $I_{OH} = O\otimes I_{PSH}$. In conventional FINCH, i.e., as a digital holography system, at least three camera shots with three different phase shifts are recorded and combined to form a complex hologram. This complex hologram when numerically back propagated, reconstructs the object information without the twin image and bias terms. In FINCH, the super resolution is not originated from the phase-shifting but from the recording configuration. So, the super resolution features are present in every FINCH hologram recorded under optimal conditions. Now applying the concepts of CAI, the image of the object can be reconstructed by a correlation. The reconstructed image ($I_R$) is given as $I_R = I_{OH} * I_{PSH}$, where '$*$' is a 2D correlation operator. Substituting for $I_{OH}$ in the above equation, we obtain $I_R = O\otimes I_{PSH} * I_{PSH}$. If $I_{PSH} * I_{PSH}$ is a Delta-like function, then the expression is simplified further as $I_R = O\otimes\delta$. Instead of correlation, in this study LR$^2$A which has been proved to generate sharpest Delta-like functions for deterministic optical fields [28]. The $I_R$ with LR$^2$A is given as $I_R = I_{OH} \odot I_{PSH}$, where '$\odot$' represents the LR$^2$A operator with optimal values of $\alpha$, $\beta$ and $p$.

In the previous studies, the resolution limit of FINCH was affected by the size of the pinhole. In this study, the $I_{PSH}$ is engineered as follows. A $I_{PSH}$ is recorded for a pinhole and used as $I_{OH}$ along with the direct imaging result $O$ of the pinhole to estimate the ideal $I_{PSH}$. The ideal $I_{PSH}$ is given as $I_{PSH} = I_{OH} \odot I_D$, where $I_D$ is the direct image of the pinhole. The aberration theory and related corrections are applied mostly with direct imaging concepts and not with holography techniques. The main reason for not needing aberration concepts and corrections in many holography systems is that the phase shifting procedure besides removing the twin image and bias terms also removes the aberrations that are common in the recorded holograms. This can be understood from the reconstruction mechanism used for conventional FINCH. The FINCH holograms recorded are not with ideal optical systems but reconstructed numerically with an ideal quadratic phase function. This is possible as the phase shifting procedure removes the aberrations present in the FINCH recording system. When implementing FINCH in the framework of CAI, the aberration function is included in the recorded $I_{PSH}$. During reconstruction by cross-correlation or LR$^2$A, the image of the object is reconstructed without the aberration. Therefore, it will be possible to use a synthetic ideal $I_{PSH}$ generated in the computer to reconstruct the $I_{OH}$ if the system is devoid of aberrations. It must be noted that this is not a special condition but the condition of most direct imaging systems. By constructing a FINCH recording setup that has minimum aberrations will enable the use of ideal synthetic $I_{PSH}$ for reconstructing FINCH holograms in the framework of CAI. The other CAI techniques such as CAFIR [30] and SWITSH [31] require changing the original optical configuration of FINCH.

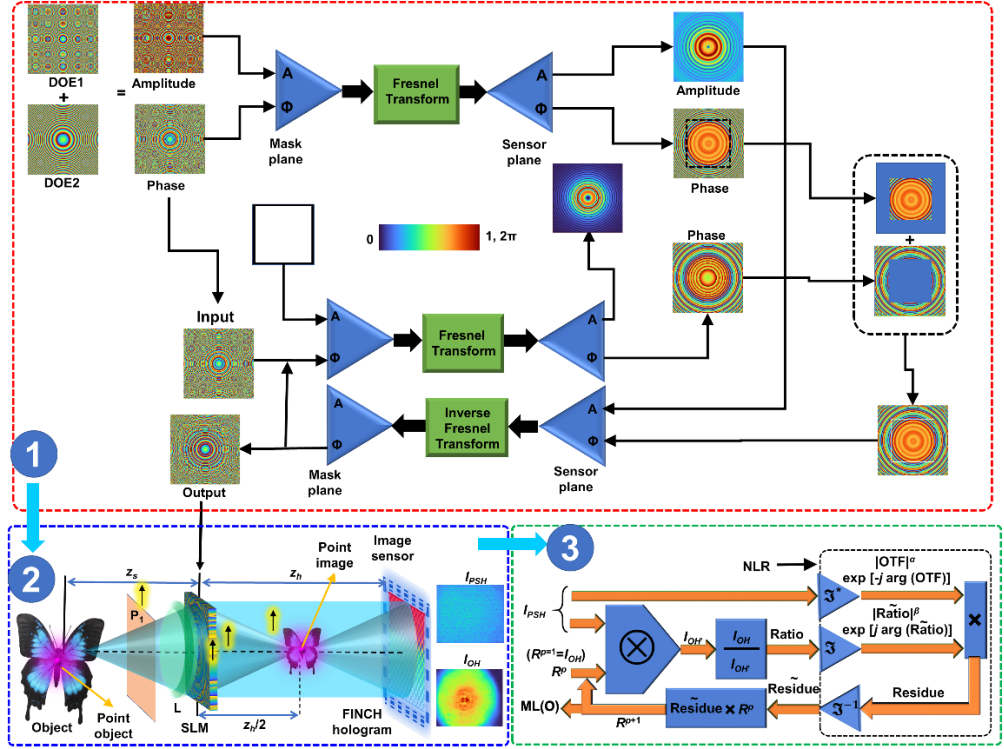

**Figure 1**. Computational optical configuration of FINCH as a CAI system. The technique consists of three major steps: computational mask design using TAP-GSA, optical FINCH experiment and recording of PSH and OH and computational reconstruction using LR$^2$A. OTF—Optical transfer function; $p$—number of iterations; $\otimes$—2D convolutional operator; $\mathfrak{J}^*$—refers to complex conjugate following a Fourier transform; $\mathfrak{J}^{-1}$- Inverse Fourier transform; $R^p$ is the $p^{th}$ solution and $p$ is an integer, when $p = 1$, $R^p = I_{OH}$; ML – Maximum Likelihood; $\alpha$ and $\beta$ are varied from -1 to 1.

## 3. Simulation studies

The simulation study of FINCH in the framework of CAI was carried out using MATLAB. A matrix size of 500 × 500 pixels was used with a pixel size $\Delta = 10$ μm and wavelength $\lambda = 650$ nm. The object distance $z_s = 50$ cm, recording distance $z_h = 1$ m and the focal lengths of diffractive lenses are 25 cm and 50 cm. The logo of University of Tartu was used as a test object for the simulation studies as shown in Figure 2(a). In conventional FINCH, at least three holograms $I_{OH1}$, $I_{OH2}$ and $I_{OH3}$ are recorded with phase shifts $\theta = 0$, $2\pi/3$ and $4\pi/3$ respectively and combined as $H_C = I_{OH1}(\exp[-i4\pi/3] - \exp[-i2\pi/3]) + I_{OH2}(1 - \exp[-i4\pi/3]) + I_{OH3}(\exp[-i2\pi/3] - 1)$. The complex hologram $H_C$ is propagated numerically to the plane of the object by $I_R = H_C \otimes Q(1/z_R)$, where $z_R$ is the reconstruction distance which is 50 cm in this case. The images of $I_{OH1}$, $I_{OH2}$ and $I_{OH3}$ are shown in Figures 2(b)-2(d) respectively. The magnitude and phase of $H_C$ are shown in Figures 2(e) and 2(f) respectively. The reconstructed image $I_R$ is shown in Figure 2(g). The direct imaging result is shown in Figure 2(h). Comparing Figures 2(g) and 2(h) an improvement in lateral resolution can be observed for FINCH.

When FINCH is implemented as CAI, there are only two holograms needed: $I_{PSH}$ and $I_{OH}$. The image is reconstructed by processing the above holograms using LR$^2$A. But the phase mask is designed using TAP-GSA. The image of the phase mask designed with 100% degrees of freedom (DoF) is shown in Figure 2(i). The $I_{PSH}$ and $I_{OH}$ are shown in Figures 2(j) and 2(k) respectively. The reconstructed image using LR$^2$A is shown in Figure 2(l). Comparing Figure

2(l) with Figures 2(g) and 2(h), an improvement in lateral resolution with respect to 2(h) and a higher SNR compared to 2(g) are seen. The structural similarity index (SSIM) values for FINCH as CAI, direct imaging and FINCH with three shots are 0.7038, 0.6529 and 0.0417 respectively. The simulation study shows a higher SNR quantified by SSIM in FINCH as CAI compared to both direct imaging and conventional FINCH. However, the lateral resolutions of Figure 2(g) and 2(l) are the same.

The axial characteristics are investigated next. A point object is considered which is shifted from $z_s$ = 30 cm to 70 cm in steps of 4 mm and the FINCH hologram is simulated and reconstructed with the reconstructing function corresponding to $z_s$ = 25 cm. In the case of conventional FINCH, the reconstruction distance was maintained the same at $z_R$ = 25 cm and in the case of FINCH as CAI, the $I_{PSH}$ corresponding to $z_s$ = 25 cm is used. The reconstructed image's intensity value at the optical axis ($x=y=0$) is plotted for direct imaging, conventional FINCH and FINCH as CAI as shown in Figure 3. It can be seen from the plots that FINCH when implemented as CAI has a higher axial resolution than conventional FINCH and the same axial resolution as direct imaging system. So unlike CAFIR, where the axial resolution of COACH and lateral resolution of FINCH was obtained, in the proposed approach, FINCH as CAI, the above high axial and lateral resolution has been achieved without any modification to the phase mask.

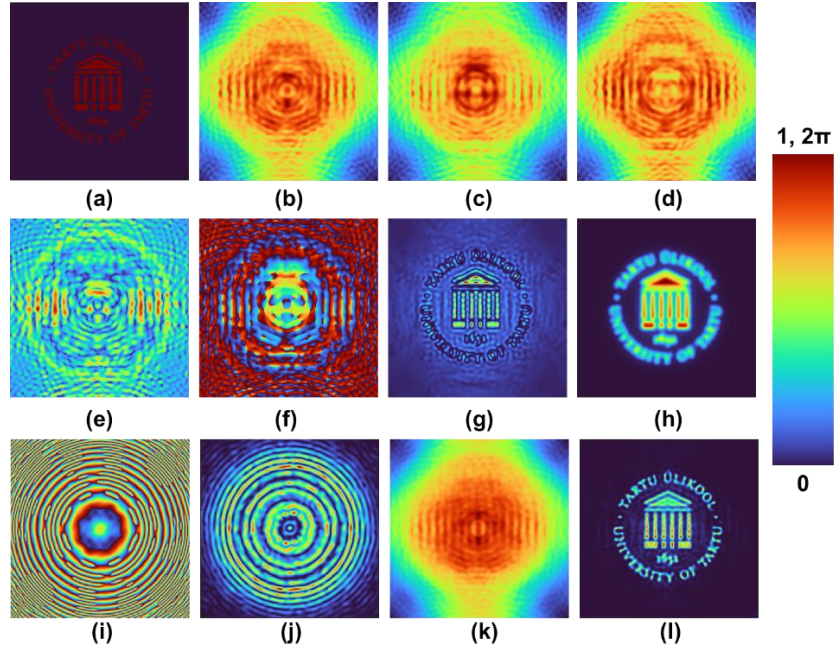

**Figure 2.** Simulation results of conventional FINCH. (a) Test object. FINCH holograms simulated for phase shifts $\theta$ = (b) 0, (c) $2\pi/3$ and (d) $4\pi/3$. (e) Magnitude and (f) phase of the complex hologram $H_C$. (g) Reconstruction result. (h) Direct imaging result. (i) Phase image of the mask designed using TAP-GSA with 100% DoF. Images of (j) $I_{PSH}$, (k) $I_{OH}$ and (l) $I_R$.

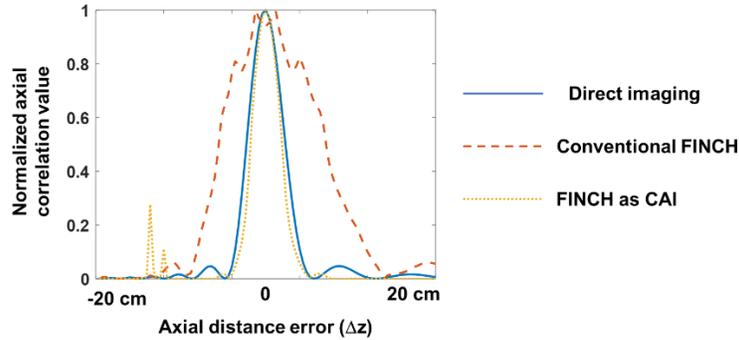

**Figure 3.** Plots of axial correlation curves for direct imaging, conventional FINCH and FINCH as CAI.

The primary lateral resolution limit in CAI methods is imposed by the numerical aperture (NA) of the system given as $\sim\lambda/NA$. As long as the diameter $d$ of the pinhole used for recording $I_{PSH}$ is smaller than $\lambda/NA$, there is no secondary resolution limit. When $d > \lambda/NA$, the pinhole samples the object with a low resolution. In a typical experiment, the lateral resolution limit imposed by NA is 1 to 10 μm which demands a high photon budget to bring the intensity distribution within the dynamic range of the camera. Even if the above is achieved, the noise level in the detector is same as the signal level resulting in a low SNR. Therefore, in many CAI techniques [26-28], some resolution was often sacrificed to achieve a high SNR. In FINCH as CAI, we investigate the possibilities of synthesizing an ideal $I_{PSH}$ from $I_{PSH}$ recorded for pinhole with a large $d$. A two-step reconstruction procedure is implemented: in the first step, the ideal $I_{PSH}$ is synthesized from recorded $I_{PSH}$ and direct image of the pinhole and the ideal $I_{PSH}$ is used to reconstruct the object's image from $I_{OH}$. In the configuration used for simulation, the diffraction limited spot size is 160 μm and so $d$ was set as 160 μm. An USAF 1951 object was used as the test object. The image of the test object, $I_{PSH}$, $I_{OH}$ and reconstructed result from LR$^2$A are shown in Figures 4(a)-4(d) respectively. The images of the $I_{PSH}$ for $d = 400$ μm and the corresponding reconstruction result are shown in Figures 4(e) and 4(f) respectively. The images of the engineered ideal $I_{PSH}$s obtained from Figure 4(e) is shown in Figure 4(g). Comparing Figures 4(b), 4(e) and 4(g), a better match is seen between Figures 4(b) and 4(g) than between Figures 4(b) and 4(e). The reconstruction results of LR$^2$A using the ideal $I_{PSH}$ shown in Figure 4(g) is shown in Figure 4(h). As it is seen, the image recovered from the ideal $I_{PSH}$ has a better match with the reference image compared to the image recovered from the $I_{PSH}$ obtained using a large pinhole. There are some loss of information and distortion but the higher resolution inherent in FINCH is preserved.

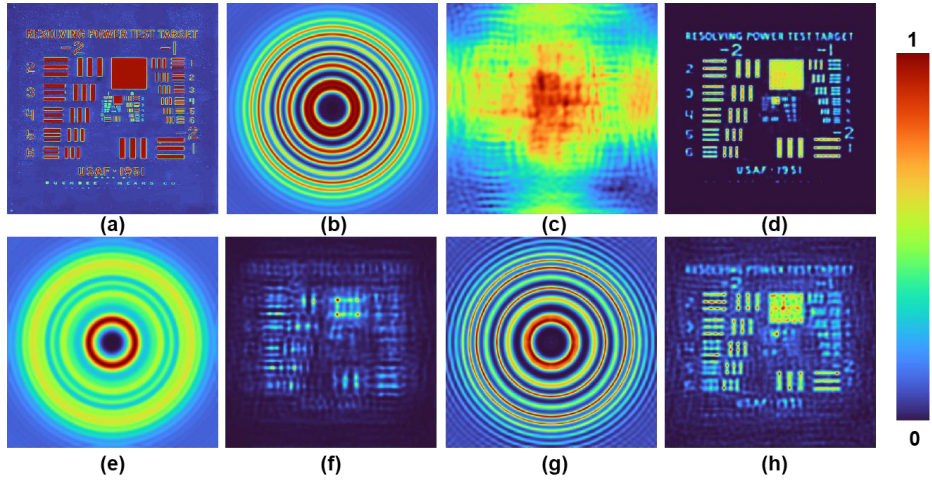

**Figure 4.** (a) Test object and images of (b) ideal $I_{PSH}$, (c) $I_{OH}$, (d) reconstruction. Images of (e) $I_{PSH}$ obtained for a larger pinhole and (f) image reconstructed using (e). (g) Image of the synthesized ideal $I_{PSH}$ and (h) its reconstruction.

## 4. Experiments

The photograph of the FINCH experimental setup is shown in Figure 5. The FINCH setup consists of the following components: high-power LED (Thorlabs, 940 mW, λ = 660 nm and Δλ = 20 nm), iris, diffuser (Thorlabs Ø1" Ground Glass Diffuser-220 GRIT), polarizer, refractive lenses, object/pinhole, beam splitter, spatial light modulator (SLM) (Thorlabs Exulus HD2, 1920 × 1200 pixels, pixel size = 8 µm) and an image sensor (Zelux CS165MU/M 1.6 MP monochrome CMOS camera, 1440 × 1080 pixels with pixel size ~3.5 µm). The light emitted from the high-power LED entered the set up through the iris which was used to control the diameter of the incoming light. Next, a diffuser was used to remove the grating lines of the LED from the incoming light. A refractive lens (L1) with a focal length of 10 cm was used to collimate the light from the diffuser. A polarizer was used to polarize the light along the active axis of the SLM. The light from the polarizer entered another refractive lens (L2) with a focal length 5 cm, which critically illuminated the object. The light from the object was collimated using another refractive lens (L3) with a focal length 5 cm and it was placed at a distance of $z_s$ = 5 cm from the object. The collimated light entered the beam splitter and reached the SLM. The phase shifted masks synthesized using TAP-GSA algorithm were displayed on the SLM and the FINCH holograms were recorded by the image sensor placed at a distance of $z_h$ = 21 cm from the SLM. The objects digit '1' and '3', were used from Group – 5 from R1DS1N—Negative 1951 USAF Test Target, Ø1" for demonstration. The $I_{PSH}$ was recorded using a 25 µm pinhole.

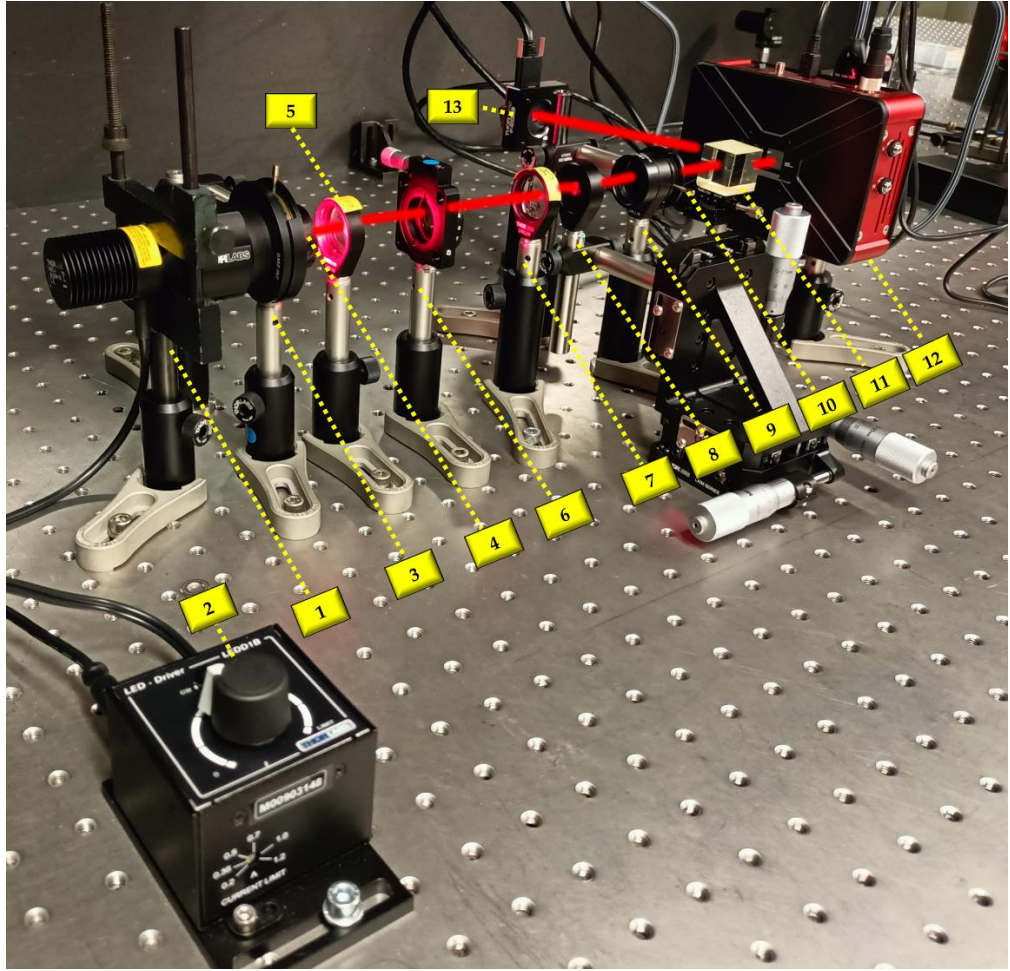

**Figure 5.** Photograph of FINCH experimental setup: (1) High-power LED, (2) LED power controller, (3) iris, (4) diffuser, (5) refractive lens L1, (6) polarizer, (7) refractive lens L2, (8) object/pinhole, (9) refractive lens L3, (10) iris, (11) beam splitter, (12) spatial light modulator, (13) image sensor. (The red line shows the path of the light beam)

Multiple experiments were carried out in the set up. In the first experiment, single plane imaging was carried out to compare the SNR and RMSE between conventional FINCH with three camera shots and FINCH based on CAI. The digit '1' from Group – 5 was used for this study. The TAP-GSA was run with a DoF of 98%. The images of the phase masks designed using TAP-GSA for $\theta = 0$, $2\pi/3$ and $4\pi/3$ are shown in Figures 6(a)-6(c) respectively. The recorded FINCH holograms $I_{OH}$s for the object digit '1' for $\theta = 0$, $2\pi/3$ and $4\pi/3$ are shown in Figures 6(d)-6(f) respectively. The above three holograms were combined to form a complex hologram whose magnitude and phase are shown in Figures 6(g) and 6(h) respectively. The reconstructed image using numerical back propagation by a distance of approximately 10 cm for conventional FINCH is shown in Figure 6(i). The FINCH hologram $I_{PSH}$s for a pinhole with a diameter of 25 μm for $\theta = 2\pi/3$ is shown in Figure 6(j). The reconstruction result by LR$^2$A for ($\alpha = 0.1$, $\beta = 1$, $p = 9$) FINCH as CAI is shown in Figure 6(k). The direct imaging result is shown in Figure 6(l). The structural similarity index (SSIM) and root mean squared error (RMSE) values were calculated using direct imaging result as reference and plotted in Figures 6(m) and 6(n) respectively. The SSIM values obtained for conventional FINCH and FINCH as CAI are 0.6668 and 0.9847 respectively. The RMSE values obtained for conventional FINCH and FINCH as CAI are 0.0278 and 0.0492 respectively. By SSIM as figure of merit, FINCH as CAI using LR$^2$A has a significantly better performance in comparison to conventional FINCH. By RMSE as figure of merit, FINCH as CAI has nearly the same performance as conventional FINCH. The above two comparison results indicate a better performance of the proposed method over the conventional method. During the study, it was noticed that conventional FINCH does not give an optimal result for all values of DoF and so different masks were tried for conventional FINCH whereas FINCH as CAI generated consistent result for different values of DoF. In the rest of the manuscript, only the optimal results of conventional FINCH were shown but FINCH as CAI experiment was carried out with the phase mask with DoF = 98%.

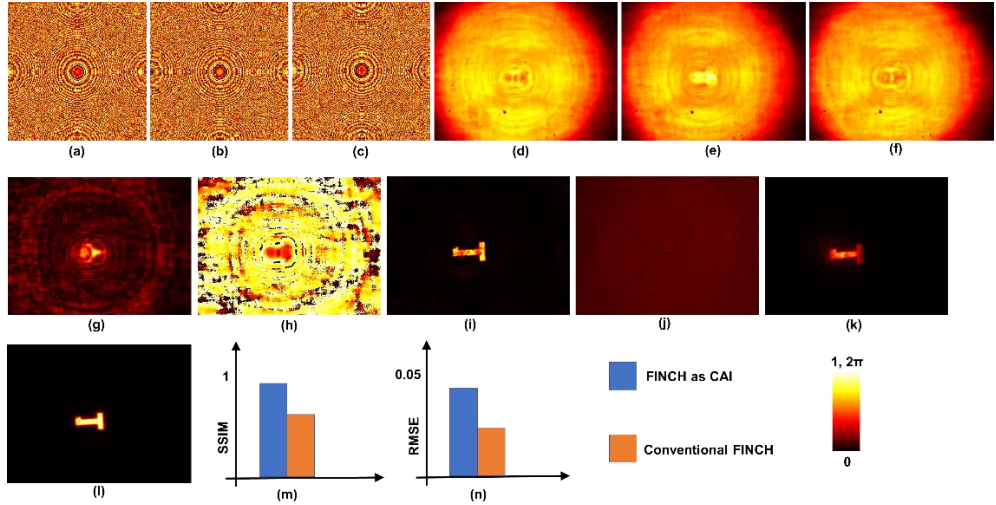

**Figure 6**. Images of the phase masks designed using TAP-GSA with phase shifts (a) $\theta = 0$, (b) $\theta = 2\pi/3$ and (c) $\theta = 4\pi/3$. $I_{OH}$ recorded using phase masks with phase shifts (d) $\theta = 0$, (e) $\theta = 2\pi/3$ and (f) $\theta = 4\pi/3$. (g) Magnitude and (h) phase of the complex hologram. (i) Reconstruction result of conventional FINCH. (j) $I_{PSH}$ recorded for $\theta = 2\pi/3$. (k) Reconstruction result of FINCH as CAI. (l) Direct imaging result. (m) Plot of SSIM for conventional FINCH and FINCH as CAI. (n) Plot of RMSE for conventional FINCH and FINCH as CAI.

In the next experiment, the axial resolution of FINCH as CAI was compared with conventional FINCH by a 3D imaging experiment. Two test objects digit '1' and digit '3' were mounted at two different distances of $z_s = 5$ and 5.6 cm. Conventional FINCH experiment was carried out using phase masks designed with DoF of 56%. The FINCH holograms $I_{OH}$s were recorded and summed to obtain the holograms for the two objects. The images of the FINCH holograms $I_{OH}$s, magnitude and phase of the complex hologram and reconstructed result are shown in Figures 7(a)-(f) respectively. The image of the object was reconstructed at a distance of approximately 10 cm. The images of the $I_{PSH}$s recorded for $z_s = 5$ and 5.6 cm are shown in Figures 7(g) and 7(h) respectively. The FINCH hologram $I_{OH}$ is shown in Figure 7(i). The reconstruction results using $I_{PSH}$ for $z_s = 5$ and 5.6 cm are shown in Figures 7(j) and 7(k) respectively. By comparing the results of Figures 7(j) and 7(k), an improvement in axial resolution in FINCH as CAI can be seen in comparison to conventional FINCH. The reconstruction parameters for LR$^2$A are same as the previous experiment.

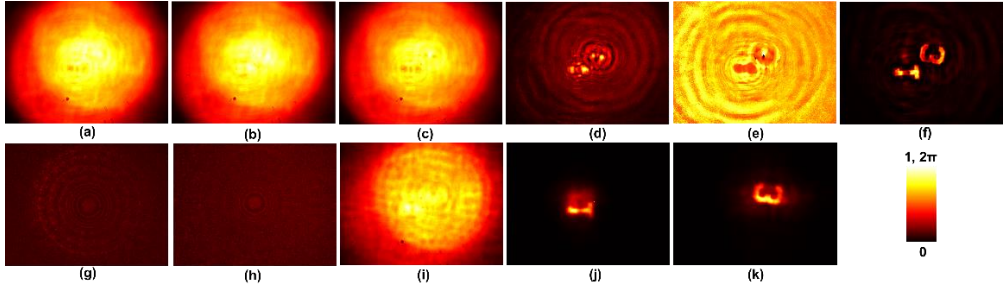

**Figure 7**. $I_{OH}$ recorded using phase masks with phase shifts (a) $\theta = 0$, (b) $\theta = 2\pi/3$ and (c) $\theta = 4\pi/3$. (d) Magnitude and (e) phase of the complex hologram. (f) Reconstruction result of conventional FINCH. (g) $I_{PSH}$ recorded for $z_s = 5$ cm. (h) $I_{PSH}$ recorded for $z_s = 5.6$ cm. (i) $I_{OH}$ for the object with two planes. (j) Reconstruction result using $I_{PSH}$ recorded for $z_s = 5$ cm. (k) Reconstruction result using $I_{PSH}$ recorded for $z_s = 5.6$ cm.

Next, we investigate the possibility of obtaining an ideal $I_{PSH}$ from the recorded $I_{PSH}$ with a pinhole with $d = 50$ μm. The image of the recorded $I_{PSH}$ is shown in Figure 8(a). The ideal $I_{PSH}$ was engineered from the recorded $I_{PSH}$ using the direct image of the pinhole. Alternatively, it can be also be just a disc created in an empty matrix with the size of the active region of camera. The image of the synthesized ideal $I_{PSH}$ is shown in Figure 8(b). The LR$^2$A was operated with ($\alpha = 0.4, \beta = 1, p = 10$). The image of the $I_{OH}$ is shown in Figure 8(c). The reconstructed results using 8(a) and 8(b) are shown in Figures 8(d) and 8(e) respectively. The LR$^2$A was operated with ($\alpha = 0.2, \beta = 1, p = 5$). The direct imaging result is shown in Figure 8(f). The reconstructed result shown in Figure 8(d) has a low resolution due to the large value of $d$ but the reconstructed result shown in Figure 8(e) has a higher sharpness similar to the results showed in Figures 6(k) and 7(j) respectively. The SSIM values of Figures 8(d) and 8(e) are 0.9766 and 0.9782 respectively. The RMSE values of Figures 8(d) and 8(e) are 0.0649 and 0.0634 respectively.

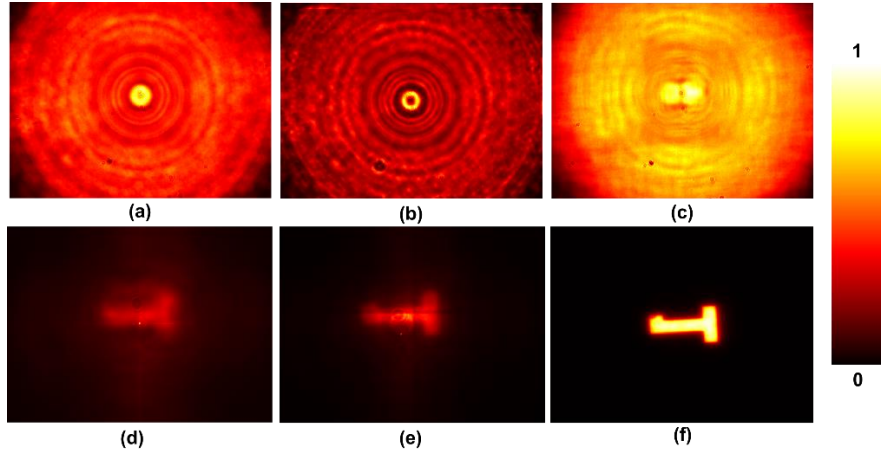

**Figure 8**. (a) $I_{PSH}$ recorded using a pinhole with $d = 50$ μm. (b) Engineered ideal $I_{PSH}$. (c) $I_{OH}$. (d) Reconstruction result using recorded $I_{PSH}$ shown in (a). (e) Reconstruction result using recorded $I_{PSH}$ shown in (b). (f) Direct imaging result.

Finally, we verify if the presented approach using LR$^2$A and TAP-GSA preserves the high axial resolution inherent to FINCH. To investigate this, an experiment was carried out using a pinhole with $d = 25$ μm which was shifted by the least count of the screw gauge which is 10 μm along $x$ and $y$ directions and FINCH holograms were recorded for each location and then summed. An aperture was added to the SLM to block light beyond a radius of 75 pixels. The results of direct imaging, conventional FINCH and FINCH as CAI are shown in Figures 9(a)-9(c) respectively. Comparing the figures, the inherent higher resolution in FINCH is seen in both 9(b) and 9(c). This validates that the developed technique preserves the high lateral resolution of FINCH.

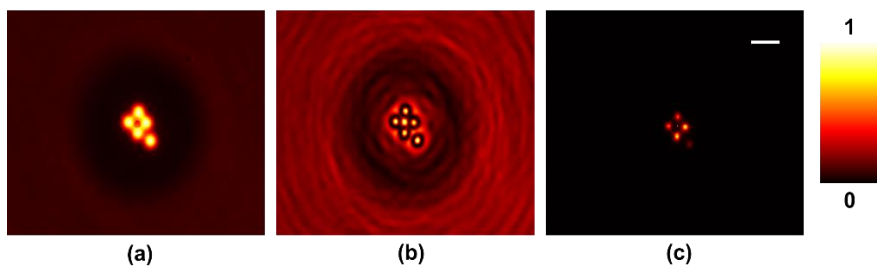

**Figure 9**. Reconstruction results of (a) direct imaging, (b) conventional FINCH and (c) FINCH as CAI. The line has a length of 100 μm.

## 5. Summary and conclusion

In this study, FINCH has been implemented as CAI by recording $I_{PSH}$ in addition to a single $I_{OH}$ and reconstructed using $LR^2A$ instead of three camera shots and numerical back propagation as in conventional FINCH. Further, TAP-GSA has been used to improve the efficiency of the the FINCH system. This approach of FINCH as CAI using $LR^2A$ and TAP-GSA enhances the imaging characteristics of FINCH in axial resolution, SNR and efficiency while preserving its inherent high lateral resolution. In this study, an inherent problem of CAI in recording $I_{PSH}$ was also addressed by a PSF engineering approach where by using a recorded $I_{PSH}$ and recorded direct image of the pinhole, the ideal $I_{PSH}$ has been synthesized. This approach prevents the loss of resolution due to $I_{PSH}$ recorded using a larger pinhole. The preliminary results are promising for extending the developed concepts for implementing FINCH to fluorescence microscopy. There are other conditions for which the method needs to be tested. In simulation, there was no constraint on the field of view as seen in the simulation results. However, in experiments, the field of view has an influence on the SNR. Further studies are needed to understand the performance of FINCH as CAI for different imaging conditions.

**Funding.** This research was funded by European Union's Horizon 2020 research and innovation programme grant agreement No. 857627 (CIPHR).

**Acknowledgments.** We thank Etienne Brasselet, CNRS, University of Bordeaux, France, for his support regarding the optical experiments.

**Data availability.** The data can be obtained from the authors upon reasonable request.

**Conflict of interests.** The authors declare no conflict of interests.

**Author Contributions:** Conceptualization, V. A.; methodology, V. A., S. J., A. P. I. X., F. G. A., S. G., A. S. J. F. R.; software, V. A., S. J.; validation, V. A., A. P. I. X., F. G. A., S. G., A. S. J. F. R.; investigation, all the authors; resources, V.A., S. J.; writing—original draft preparation, A. P. I. X., F. G. A., V. A.; writing—review and editing, all the authors; supervision, V. A., S. J.; project administration, A. S. J. F. R.; funding acquisition, V.A., S. J. All authors have read and agreed to the submitted version of the manuscript.